\documentstyle[titlepage,fleqn,12pt]{article}
\newcommand{\be}{\begin{equation}}
\newcommand{\ee}{\end{equation}}
\newcommand{\bea}{\begin{eqnarray}}
\newcommand{\eea}{\end{eqnarray}}
\newcommand{\bdm}{\begin{displaymath}}
\newcommand{\edm}{\end{displaymath}}
\begin{document}
\begin{titlepage}
\title{On bosonization in $3$ dimensions.}
\author{D. G. Barci
\footnotemark[1]\\
C. D. Fosco \\
and L. E. Oxman
\footnotemark[2]
\\ \\ High Energy Group\\
International Centre for Theoretical Physics\\
P.O. Box 586 \\34100 Trieste \\Italy.}
\date{}
\footnotetext[1]{Permanent address:
Instituto de F\'\i sica (FIT) , Universidade Federal do Rio de
Janeiro C.P. 68528, Rio de Janeiro, RJ, 21945-970, Brasil}
\footnotetext[2]{Permanent address:
Departamento de F\'\i sica, Facultad de Ciencias Exactas
y Naturales,
Universidad de Buenos Aires, Ciudad Universitaria, 1428, Buenos
Aires, Argentina}
\vspace{1cm}
\baselineskip=21.5pt
\begin{abstract}
A recently proposed  path-integral bosonization scheme for massive
fermions in $3$ dimensions is extended by keeping the full momentum-dependence
of the one-loop vacuum polarization tensor. This makes it possible
to discuss both the massive and massless fermion cases on an equal
footing, and moreover the results it yields for massless
fermions are consistent with the ones of another, seemingly different,
canonical quantization approach to the problem of bosonization for
a massless fermionic field in $3$ dimensions.
\vskip 2cm
Key words: Bosonization, $QED_3$.
\vskip 0.7cm
Submitted to: Physics Letters B.
\end{abstract}
\maketitle
\end{titlepage}
\baselineskip=21.5pt
\parskip=3pt

During the last few years many different proposals have been considered
to bosonize fermionic theories in $3$
dimensions~\cite{haldane,marino,kovner,bur1,bur2,frascha,fidel}.
In Ref.~\cite{marino},
order-disorder field operators related to a free massless Dirac field were
defined. Applying canonical
quantization methods, a bosonic, non-local and gauge-invariant action
for an Abelian vector field was constructed, the approximate
bosonization rules (in Euclidean spacetime) being
\bea
\bar{\psi}\not \! \partial \psi &\leftrightarrow&
\frac{1}{4}\, F_{\mu\nu}(-\partial^2)^{-1/2}F_{\mu\nu}
\,+\,\frac{i}{2}\,\theta \,\epsilon_{\mu\nu\lambda}A_\mu\partial_\nu
A_\lambda \,+\, nqt \nonumber \\
\bar{\psi}\gamma_\mu\psi &\leftrightarrow&
\beta \, \epsilon_{\mu\nu\lambda}\partial_\nu A_\lambda
\,-\,\beta \,\theta \,(-\partial^2)^{-1/2} \partial_\nu F_{\mu\nu}+ nqt
\label{1.1}
\eea
where $\psi$ is a two-component Dirac spinor, $A_\mu$ is a $U(1)$
gauge field, and $nqt$ means non-quadratic terms in $A_\mu$ (the
neglecting of non-quadratic terms is what makes this bosonization
approximate). The parameter
$\theta$ is regularization-dependent. This sort of ambiguity, which
manifests itself in the bosonization rules, already exists in
the fermionic description. It is due to the regularization
dependence of the induced Chern-Simons term~\cite{regdep}.

In Ref.~\cite{frascha}, functional methods were applied
to derive bosonization formulae for the free massive
Thirring model, and in ~\cite{fidel}, the Abelian and
non-Abelian cases in any dimension $d \geq 2$ were considered.
These `long distance' bosonization rules are
reliable for the description of phenomena where the fermionic
current is not a strongly varying field, with a typical scale
of variation much bigger that the inverse of the fermion mass.
In this regime, either the free massive Dirac field or the
Thirring model (in $3$ dimensions) can be mapped to Chern-Simons
theories by using the approximate bosonization rules
\bea
\bar{\psi}(\not\! \partial\,+\,m)\psi &\leftrightarrow&
\pm \frac{i}{2}\epsilon_{\mu\nu\lambda}A_\mu\partial_\nu A_\lambda
\nonumber \\
\bar{\psi}\gamma_\mu \psi &\leftrightarrow&
\pm i\sqrt{\frac{1}{4\pi}}\;\epsilon_{\mu\nu\lambda} \partial_\nu A_\lambda
\label{1.2}
\eea
This is valid to leading order in $\displaystyle{\frac{1}{m}}$, while the
inclusion of the next-to-leading order term would lead to a
Maxwell-Chern-Simons theory instead.

As it was stressed in ~\cite{frascha,fidel}, the possibility of finding
exact bosonization rules (in this functional approach), depends on our
ability to compute the
fermionic determinant in the presence of a background field exactly.
Thus in $3$ dimensions we must use an approximation scheme. The one
presented in \cite{frascha,fidel}
amounts to expanding the corresponding effective action in powers of
$\displaystyle{\frac{\partial}{m}}$.

The question presents itself about how to extend this approximation in
order to include cases where the derivative expansion
is no longer valid, as it is indeed the case for massless
fermions. The most obvious attempt to improve the approximation
would be to include higher order terms in the derivative expansion.
However, when doing so a new problem arises. The resulting
theory will present instabilities, which in the Euclidean formulation
are manifested in the action being not bounded from below, whereas
in Minkowski space the related unitarity problem shows up.
It is possible to get rid of this apparent drawback by recalling
that the effective higher-order theory is valid only for gauge
fields with momenta smaller than a cut-off of the order of the
fermion mass $m$, which is a region free
from such unphysical features, as can be easily verified.
At any rate, cases where the momentum is larger than the
fermion mass remains out of the scope of any (however refined) derivative
expansion.

In this letter we attempt to overcome this kind of limitation by including
the full momentum dependence in the one-loop quadratic part of the
effective action. Whence the results will also be valid for the
massless case, without spoiling the proper low-momentum features.
As no momentum expansion is performed, there is no instability
problem. Keeping the full momentum dependence one introduces a non-locality
in the bosonized action, a property
shared with the approach of \cite{marino}. This non-locality is unavoidable
as soon as the derivative expansion, which always produces local
terms, is discarded. For massless fermions in particular, one
cannot escape the non-locality, since there is a branch cut at
zero momentum so the one-loop vacuum-polarization tensor cannot be
analytic there.

The above mentioned approaches, canonical and functional, to bosonization
in $3$ dimensions look {\em a priori} quite different and their relationship
is not at all obvious. We will show that, by keeping the full momentum
dependence of the vacuum polarization tensor in the approach of \cite{fidel},
one can reproduce \cite{marino} if the mass of the
Dirac field is set equal to zero. The result of \cite{fidel} will survive
in the low-momentum (or $m \to \infty$) limit.

We start by constructing a bosonized version of the generating
functional of current correlation functions in the case of a free
fermionic field in three dimensions,
reviewing the procedure followed in \cite{fidel}. This method
builds upon the functional
representation of the fermionic generating functional
\be
Z(s)=\int [d\psi][d\bar{\psi}]\;
\exp \left[ -\int d^3x \;\bar{\psi}( \not \! \partial +i \not \! s +m)
\; \psi \right]
\label{1.3}
\ee
by performing the change of variables
\be
\psi(x)\rightarrow e^{i\alpha(x)}\,\psi(x)
\makebox[.5in]{,}
\bar{\psi}(x)\rightarrow e^{-i\alpha(x)}\,\bar{\psi}(x)
\label{1.4}
\ee
to obtain
\be
Z(s)=\int [d\psi][d\bar{\psi}] \;
\exp \left[ -\int d^3x \;\bar{\psi}(\not \!\partial
+i(\not \! s+ \not \! \partial\alpha)+m)\,\psi \right]\;.
\label{1.5}
\ee
Defining $b_{\mu}=\partial_{\mu}\alpha$ (\,$\Rightarrow \, F_{\mu \nu}(b)=
\partial_\mu b_\nu - \partial_\nu b_\mu \,=\,0$
\,), as $Z(s)$ does not depend on $b_\mu$, the pure-gauge field $b_\mu$ can be
integrated with an arbitrary (non-singular) weight functional $f(b)$, yielding
(up to a normalization factor)
\begin{eqnarray}
Z(s) = \int [db][d\psi][d\bar{\psi}] &f(b)&
\delta(F_{\mu \nu}(b)) \,
\exp  -\int d^3x \bar{\psi}(\not \! \partial
+i(\not \! s+ \not \! b)+m)\psi \nonumber\\
&=& \int [db][d\psi][d\bar{\psi}] f(b-s)
\delta(F_{\mu \nu}(b - s)) \nonumber\\
&\times& \exp -\int d^3x \bar{\psi}  (\not \! \partial
+i \not \! b  +  m) \psi \,,
\label{1.6}
\end{eqnarray}
where the last equation follows from the first one by shifting
$b \, \to \, b - s$.
Introducing a Lagrange multiplier $A_{\mu}$ to
exponentiate the $\delta$-functional, integrating over the fermion
fields and setting the weight functional equal to one, it yields
\be
Z(s)=\int [dA]\, [db]\, \exp \left[ -T(b)\,-\,i \, \int d^3x \, A_{\mu}
(\epsilon_{\mu \nu \lambda} \partial_\nu b_\lambda \,-\,
\epsilon_{\mu \nu \lambda} \partial_\nu s_\lambda) \right]
\label{1.7}
\ee
where $T (b)$ denotes the fermionic effective action in the presence of
an external vector field
\be
T(b) \; = \; -\log \det (\, \not \! \partial \, + \,i\, \not \! b \, + \, m \,)
\;.
\label{1.8}
\ee
We now make the approximation of retaining up to quadratic terms in $b_\mu$
in (\ref{1.8}). This is consistent with the approaches of ref.'s \cite{marino}
and \cite{fidel} \footnote{This is equivalent to introducing a
`coupling constant' $e$ by means of the redefinition
$b_\mu \to e \, b_\mu$, and working up to order in $e^2$.}.
The quadratic part of $T(b)$ may be split as
\bea
T(b) \,&=&\,  T_{PC}(b) \; + \; T_{PV}(b) \nonumber \\
T_{PC}(b)\,&=& \,\int d^3 x \, \frac{1}{4} \,
F_{\mu \nu}(b)\, F(-\partial^2) \; F_{\mu \nu}(b)\nonumber \\
T_{PV}(b)\,&=&\,\int d^3 x \, \frac{i}{2} \, b_{\mu} \, G(-\partial^2)
\, \epsilon_{\mu \nu \lambda}\partial_{\nu}b_{\lambda}
\label{1.9}
\eea
where $T_{PC}$ and $T_{PV}$ come from the parity-conserving and
parity-violating
pieces of the vacuum-polarization tensor, respectively~\cite{loop}.
The function $F$ in (\ref{1.9}) is regularization-independent, and a
standard one-loop calculation yields
\be
{\tilde F} (k^2) \;=\; \frac{\mid m \mid}{4 \pi k^2} \,
\left[ 1 - \displaystyle{\frac{1 \,-\,\displaystyle{\frac{k^2}{4 m^2}}}{(
\displaystyle{\frac{k^2}{4 m^2}})^{\frac{1}{2}}}} \, \arcsin(1\,+
\, \frac{4 m^2}{k^2})^{-\frac{1}{2}} \right] \;,
\label{1.10}
\ee
where here and in what follows we shall always denote momentum-space
representation by putting a tilde over the corresponding coordinate-space
representation quantity.
The function ${\tilde G}$ in (\ref{1.9}) is regularization {\em dependent}, and
can be written as
\be
{\tilde G} (k^2) \;=\; \frac{q}{4 \pi} \,+\, \frac{m}{2 \pi \mid k \mid}
\, \arcsin (1 \, + \, \frac{4 m^2}{k^2} )^{- \frac{1}{2}} \;,
\label{1.11}
\ee
where $q$ can assume any integer value~\cite{regdep,regdep1}, and may be
thought of
as the effective number of Pauli-Villars regulators, namely, the
number of regulators with positive mass minus the number of
negative mass ones.
Adding a gauge-fixing term $\displaystyle{\frac{\lambda}{2}}
(\partial \cdot b)^2$, the $b$ - dependent
part of the path integral (in momentum-space) reads:
\be
 I = \int [db] e^{\displaystyle{{-\frac{1}{2}\,\int \,\frac{d^3 k}{(2\pi)^3} \,
( {\tilde b}^{\dagger}(k) \,{\tilde M}(k)\, {\tilde b}(k) + i
{\tilde b}^{\dagger}(k) |k| \, ({\tilde P}_+ (k) - {\tilde P}_- (k) )\,
{\tilde A}(k)) }}} \,.
\label{1.12}
\ee
We introduced an obvious matrix notation,
where the fields are represented by column vectors, the matrix ${\tilde M}$ is
given by
\be
{\tilde M}(k)\; = \; ( {\tilde F} k^2 \,+\, i \,{\tilde G} \mid k \mid )
\, {\tilde P}_+ \,+\, ( {\tilde F} k^2 \,-\, i \,{\tilde G} \mid k \mid ) \,
{\tilde P}_- \,+\,  \lambda k^2 \, {\tilde L} \; ,
\label{1.13}
\ee
and we introduced a complete set of hermitian orthogonal projectors
\be
({\tilde P}_{\pm})_{\mu \nu}\,=\,\frac{1}{2} \, \left(\,\delta_{\mu \nu}
- \frac{k_{\mu} k_{\nu}}{k^2} \,\pm \, i
\epsilon_{\mu \lambda \nu} \frac{k_{\lambda}}{\mid k \mid}\, \right)
\;\;,\;\;
{\tilde L}_{\mu \nu} \,=\, \frac{k_{\mu} k_{\nu}}{k^2} \;,
\label{1.14}
\ee
which verify ${\tilde P}_{\pm}^2={\tilde P}_{\pm}$, ${\tilde L}^2={\tilde L}$;
${\tilde P}_{\pm}{\tilde L}=0$, ${\tilde P}_+ {\tilde P}_-=0$;
and ${\tilde P}_+ + {\tilde P}_- + {\tilde L} = 1$.

The bosonization formulae are obtained by integrating
out the ${\tilde b}$-field
\be
\!\!\! I \,=\, \exp \left[ - \frac{1}{2}\,\int \, \frac{d^3 k}{(2\pi)^3}\,
{\tilde A}^{\dagger}\,({\tilde P}_+ (k) - {\tilde P}_- (k)) \,k^2 \,
{\tilde M}^{-1} ({\tilde P}_+ (k) - {\tilde P}_- (k)) \, {\tilde A} (k)
\right]
\label{1.15}
\ee

The inverse of ${\tilde M}$, needed in (\ref{1.15}) is computed from
(\ref{1.13}),
\be
{\tilde M}^{-1} (k)\; = \; ( {\tilde F} k^2 \,+\, i \,{\tilde G} \mid k
\mid )^{-1} \, {\tilde P}_+ \,+\, ( {\tilde F} k^2 \,-\, i \,
{\tilde G} \mid k \mid )^{-1} \, {\tilde P}_-
\,+\, (\lambda k^2 )^{-1} \, {\tilde L} \;,
\label{1.16}
\ee
and by further use of the projectors' properties, we can write
\bea
Z(s)&=&\int [d{\tilde A}]
\exp \int  \frac{d^3 k}{(2\pi)^3}[
\frac{1}{2} k^2 {\tilde A}^{\dagger}  \nonumber\\
( \frac{1}{{\tilde F} k^2+i{\tilde G} |k|} \, {\tilde P}_+ &+&
\frac{1}{{\tilde F} k^2-i {\tilde G} |k|}\, {\tilde P}_- )  {\tilde A} \,-\,i\,
{\tilde s}^{\dagger}\, |k|\, ({\tilde P}_+ \,-\, {\tilde P}_-) \,{\tilde A} ]
\;.
\label{1.17}
\eea
There is still freedom to write the partition function (\ref{1.17}) in
different ways, namely, we can always redefine the field
${\tilde A}_{\mu}$ by performing a non-singular transformation on it.
This will, of course, change both the quadratic and linear parts
of the action, thus affecting both the bosonized action and the
mapping between fermionic currents and bosonic fields, but in
such a way that the current correlation functions are not
modified, since we are just changing a dummy variable. It is
however, necessary to do this in order to show explicitly the
connection with the approach of \cite{marino}.
A general redefinition of ${\tilde A}_{\mu}$ may be written as
${\tilde A} \, \rightarrow \, ({\tilde u}_+ \,P_+\; +\; {\tilde u}_-
P_- \;+\; {\tilde u}_L \, L ) {\tilde A}$,
where the ${\tilde u}$'s are functions of the momentum.
Note that the
effect of ${\tilde u}_L$ disappears as a consequence of gauge-invariance.
\bea
Z(s)&=&\int [d {\tilde A}]
\exp -\int \frac{d^3 k}{(2\pi)^3}
( \frac{1}{2} k^2 {\tilde A}^{\dagger} [
\frac{|{\tilde u}_+|^2}{{\tilde F}k^2+i{\tilde G}|k|}P_+ \nonumber\\
 &+& \frac{|{\tilde u}_-|^2}{{\tilde F}k^2-i{\tilde G}|k|}P_-]
{\tilde A} - i {\tilde s}^{\dagger}|k|({\tilde u}_+ P_+-{\tilde u}_-P_-)
{\tilde A})\;.
\label{1.18}
\eea
In what follows we shall restrict ourselves to the constant-${\tilde u}_\pm$
case. Expression (\ref{1.18}) can be put in coordinate space representation
as follows:
\bea
Z(s)&=&\int [dA]
\exp -\int d^3 x
[ \frac{1}{4} F_{\mu \nu} \, C_1 \, F_{\mu \nu}
- \frac{i}{2} A_{\mu} \, C_2 \, \epsilon_{\mu \nu \lambda}
\partial_{\nu} A_{\lambda}  \nonumber\\
&+& i \, (\frac{u_+ - u_-}{2}) \, s_\mu
\frac{1}{\sqrt{-\partial^2}} \partial_{\nu} F_{\nu \mu}
\,-\, i \, (\frac{u_+ + u_-}{2})\, s_\mu \epsilon_{\mu \nu \lambda}
\partial_{\nu} A_{\lambda} ]
\label{1.19}
\eea
where
\bea
C_1 &=& \frac{1}{2} \; \frac{ |u_+|^2 (F - i G) \,+\,|u_-|^2
(F + i G)}{ - \partial^2 F^2 \, + \, G^2 } \nonumber\\
C_2 &=& \frac{i}{2} \; \frac{ |u_+|^2 (F - i G) \,-\,|u_-|^2
(F + i G)}{ - \partial^2 F^2 \, + \, G^2 }
\label{1.20}
\eea

Let us discuss now the explicit form adopted by (\ref{1.20}) for the cases
$m \to \infty$ and $m \to 0$, to make contact with the results
of reference \cite{fidel} (particularized to the Abelian $d = 3$ case) and
reference \cite{marino}, respectively. This is achieved by
evaluating $C_1$ and $C_2$ in the corresponding
limits, and this is in turn determined by the values of
$F$ and $G$. When $m \to \infty$, $C_1$ tends to a constant which multiplies
the Maxwell term. This is neglected to leading order in a derivative
expansion, since there is also a Chern-Simons term,
multiplied by the constant factor $C_2$:
\be
C_2 \;\to \;\; 4 \pi \ |u|^2 \,\times \, ( q \, + \, \frac{m}{|m|} ) \;.
\label{1.21}
\ee
$C_2$ is regularization-dependent, and its ambiguity is reflected
here by the undefined constant $q$. To compare with \cite{fidel}, we
partially fix $q$ by the condition $q + {\rm sgn} (m) \, =\, \pm 1$, and
chosing $u_+ \,=\, u_- \,=\, u
\,=\, \frac{1}{2 \pi}$, we see that the bosonized action
(denoted $S_{bos}$), in the partition function (\ref{1.19}) reduces to
\begin{equation}
S_{bos} \,=\, \int d^3 x \, \left( \pm \frac{i}{2}
A_{\mu} \epsilon_{\mu \nu \lambda} \partial_{\nu} A_\lambda
\,-\, \frac{i}{\sqrt{4 \pi}} s_{\mu} \epsilon_{\mu \nu \lambda}
\partial_{\nu} A_{\lambda} \right) \;,
\label{1.22}
\end{equation}
which agrees with the result of \cite{fidel}.

Now we discuss the limit  $m\rightarrow 0$.
In this case we have for $F$ and $G$ the behaviours
\be
F(k^2) \; \to \; \frac{e^2}{16}\; |k|^{-1}
\makebox[.5in]{,}
G(k^2) \; \to \;  \displaystyle{\frac{q}{4 \pi}}
\label{1.23}
\ee
which imply for $C_1$ and $C_2$
\be
C_1 \to \frac{16 |u|^2}{|k|} \;\;\; C_2 \to \frac{4 \pi |u|^2}{q}
\;.
\label{1.24}
\ee
By taking then
\be
{\tilde u}_+ \;=\; {\tilde u}_- \;=\; \displaystyle{\frac{1}{4}} \,
e^{i \alpha} \;,
\label{1.25}
\ee
the bosonized action in coordinate space assumes the form
\bea
S_{bos} &=& \int d^3 x \, ( \frac{1}{4} \,
F_{\mu \nu} \frac{1}{\sqrt{-\partial^2}} F_{\mu \nu} \;-\;
\frac{i}{2} \, \frac{\pi}{4 q} \, \epsilon_{\mu \nu \lambda}
A_\mu \partial_\nu A_\lambda \nonumber\\
&-& \frac{ \sin \alpha }{4} \, s_\mu \, \frac{\partial_\nu
F_{\nu \mu}}{\sqrt{-\partial^2}} \;-\; i \frac{\cos \alpha}{4}
\, \epsilon_{\mu \nu \lambda} \partial_\nu A_\lambda ) \;,
\label{1.26}
\eea
thus with the identifications
\begin{equation}
\theta \;=\; \frac{\pi}{4 q} \;\;,\;\; \alpha \;=\; \arctan
\frac{\pi}{4 q} \;\;,\;\; \beta \;=\; \frac{\cos \alpha}{4}
\;,
\label{1.27}
\ee
the bosonized action becomes identical with the one of
Equation (\ref{1.1}), which is the Euclidean version of the
one of Ref.~\cite{marino}.

We have thus studied the full bosonized partition function (\ref{1.19})
for the low and large momentum regimes. In the general case, the
full expression (\ref{1.19}) should be retained. It is however, possible
to simplify the form of ${\tilde F}$ and ${\tilde G}$, by replacing them by
approximate
but simpler looking expressions, which may be replaced in
(\ref{1.20}). With an error smaller than $10$ percent
over the full range of momenta, we have the approximations:
\bea
{\tilde F} (k^2) &=& \frac{\mid m \mid}{4 \pi k^2} \,
\left[ 1 - \displaystyle{\frac{1 \,-\,\displaystyle{\frac{k^2}{4 m^2}}}{(
\displaystyle{\frac{k^2}{4 m^2}})^{\frac{1}{2}}}} \, \arcsin(1\,+
\, \frac{4 m^2}{k^2})^{-\frac{1}{2}} \right] \nonumber\\
&\simeq & \frac{1}{16} [ k^2 \,+\, ( \frac{3 \pi m}{4} )^2 ]^{-\frac{1}{2}}
\nonumber\\
{\tilde G} (k^2) &=& \frac{q}{4 \pi} \,+\, \frac{m}{2 \pi \mid k \mid}
\, \arcsin (1 \, + \, \frac{4 m^2}{k^2} )^{- \frac{1}{2}} \nonumber\\
&\simeq & \frac{q}{4 \pi} \;+\; \frac{m}{4} \, [ k^2 \, +\,
\pi^2 m^2 ]^{- \frac{1}{2}} \;,
\label{1.28}
\eea
which are obtained by following the approach of Ref.~\cite{yo}.

In this letter we have extended the method presented in
\cite{fidel}, to obtain a bosonization for the free Dirac
field, valid over the whole range of distances.
A sensible extension is achieved by retaining the complete one-loop
quadratic part of the effective action. The (non-local) bosonized theory
that is obtained in this way have some advantages with respect to the
(local) higher order theory that would have been obtained by considering a
finite number of terms, when expanding the effective action in powers
of $\displaystyle{\frac{\partial}{m}}$. On the one hand we can see that the
Euclidean action is positive definite leading to a
stable bosonized theory. On the other hand it unables us to
treat the massive and massless cases in an equal footing, leading to
the bosonization formulae for a massless Dirac field (Eq. (\ref{1.1})),
obtained by following the canonical method.

In contrast, the higher order Euclidean effective action that results
from the approximation of the non-local effective action is not
positive definite, which leads to an unstable behaviour, unless
a cut-off of the order of the fermion mass is used.

In Minkowski space, a similar situation shows up. A
higher order theory leads to the presence of poles in the field
propagator which are in conflict with unitarity. However, these poles
will be located at a mass scale greater than $m$ and again the
imposed cut-off will prevent these poles from producing unphysical
effects. Now, if we look
Minkowskian version of the non-local Lagrangian Ref.~\cite{marino}
as this equivalence is valid over the whole range
of momenta, no unphysical problems should appear. This is precisely
the case (see Ref.\cite{lucho}).

\newpage
\section*{Acknowledgements}
D. G. B. was supported by CNPq and FUJB, Brasil,
L. E. O. by CONICET , Argentina, and C. D. F.  by
ICTP, Italy.

We acknowledge F. A. Schaposnik for useful comments.


\newpage


\begin{thebibliography}{99}
\bibitem{haldane} F. D. Haldane, Helv. Phys. Acta {\bf 65} (1992) 52.
\bibitem{marino} E. C. Marino, Phys. Lett. {\bf B 263} (1991) 63.
\bibitem{kovner} A. Kovner and P. S. Kurzepa, Phys.  Lett. {\bf B 321}
(1994) 129.
\bibitem{bur1} C. P. Burgess and F. Quevedo, Nucl. Phys. {\bf B421}
(1994) 373;
C. P. Burgess and F. Quevedo, Phys. Lett. {\bf B329} (1994) 457
\bibitem{bur2} C.  Burgess, C.A. L\"utken and F. Quevedo, Phys. Lett.
{\bf B336} (1994) 18.
\bibitem{frascha} N. Brali\'c, E. Fradkin, M. V. Man\'\i as and
F. A. Schaposnik, Nucl. Phys. {\bf B 446} (1995) 144;
E. Fradkin and F. A. Schaposnik, Phys. Lett.
{\bf B 338} (1994) 253.
\bibitem{fidel} F. A. Schaposnik, `A comment on bosonization in
$d \geq 1$ dimensions', F. A. Schaposnik, hep-th/9505049, to
appear in Phys. Lett. B.
\bibitem{loop} R. Jackiw and S. Templeton, Phys. Rev. {\bf D 23} (1981) 2291;
S. Deser, R. Jackiw and S. Templeton, Phys. Rev. Lett. {\bf 48} (1982) 975;
\bibitem{regdep} J. Fr\"ohlich and T. Kerler, Nucl. Phys. {\bf B 354}
(1991) 369;
\bibitem{regdep1}`The $\zeta$-function answer to parity violation in three
dimensional gauge theories', R. E. Gamboa Saravi, G. L. Rossini and F. A.
Schaposnik,
hep-th/9411238.
\bibitem{lucho}`Canonical quantization of non-local field equations',
D. G. Barci, L. E. Oxman and M. Rocca, hep-th/9503101.
\bibitem{yo}I. J. R. Aitchison, C. D. Fosco and J. A. Zuk,
Phys. Rev. {\bf D 48} (1993) 5895.
\end{thebibliography}
\end{document}